\documentclass{aa}
\newcommand{\EQ}{\begin{equation}}
\newcommand{\EN}{\end{equation}}
\newcommand{\EQA}{\begin{eqnarray}}
\newcommand{\ENA}{\end{eqnarray}}

\newcommand{\Eq}[1]{Eq.~(\ref{#1})}

\newcommand{\Sec}[1]{Sect.~\ref{#1}}

\newcommand{\Fig}[1]{Fig.~\ref{#1}}
\newcommand{\FFig}[1]{Figure~\ref{#1}}

\newcommand{\Figs}[2]{Figs~\ref{#1} and \ref{#2}}

\newcommand{\meanA}{\overline{A}}
\newcommand{\meanB}{\overline{B}}

{}
{}
{}
{}
{}
{}
{}
\newcommand{\eee}{\hat{\mbox{\boldmath $e$}} {}}

\newcommand{\xxx}{\hat{\mbox{\boldmath $x$}} {}}

\newcommand{\xx}{\mbox{\boldmath $x$} {}}

\newcommand{\UU}{{\vec{U}}}
\newcommand{\uu}{{\vec{u}}}
\newcommand{\BB}{{\vec{B}}}
\newcommand{\JJ}{{\vec{J}}}

\newcommand{\AAA}{{\vec{A}}}

\newcommand{\ff}{\mbox{\boldmath $f$} {}}

\newcommand{\FF}{{\vec{F}}}

\newcommand{\kk}{\mbox{\boldmath $k$} {}}

\newcommand{\nab}{\vec{\nabla}}

\newcommand{\SSSS}{\mbox{\boldmath ${\sf S}$} {}}
\newcommand{\GGGG}{\mbox{\boldmath ${\sf G}$} {}}

\newcommand{\DD}{{\rm D} {}}

\newcommand{\dd}{{\rm d} {}}

\def\half{{\textstyle{1\over2}}}

\def\onethird{{\textstyle{1\over3}}}

\newcommand{\yjgr}[3]{ #1, {JGR,} {#2}, #3}

\newcommand{\ysph}[3]{ #1, {Sol. Phys.,} {#2}, #3}
\newcommand{\yapj}[3]{ #1, {ApJ,} {#2}, #3}
\newcommand{\yapjl}[3]{ #1, {ApJ,} {#2}, #3}

\newcommand{\yan}[3]{ #1, {AN,} {#2}, #3}

\newcommand{\yanas}[3]{ #1, {A\&AS,} {#2}, #3}

\newcommand{\yjfm}[3]{ #1, {JFM,} {#2}, #3}
\newcommand{\ypf}[3]{ #1, {Phys. Fluids,} {#2}, #3}
\newcommand{\ypp}[3]{ #1, {Phys. Plasmas,} {#2}, #3}

\newcommand{\yprl}[3]{ #1, {Phys. Rev. Lett.,} {#2}, #3}

\newcommand{\ypr}[3]{ #1, {Phys. Rev.} {#2}, #3}

\newcommand{\ybook}[3]{ #1, {#2} (#3)}
\newcommand{\yproc}[5]{ #1, in {#3}, ed. #4 (#5), #2}

\newcommand{\sapjl}[1]{ #1, {ApJL} (submitted)}

\newcommand{\pjour}[2]{ #1, {#2} (in press)}

\usepackage{graphicx}
\usepackage{url}
\usepackage{times}
\begin{document}
\title{Relaxation of writhe and twist of a bi-helical magnetic field}
\author{T.\ A.\ Yousef\inst{1} and A.\ Brandenburg\inst{2}}

\institute{
Faculty of Engineering Science and Technology,
Norwegian University of Science and Technology,
Kolbj{\o }rn Hejes vei 2B, N-7491 Trondheim, Norway
\and 
NORDITA, Blegdamsvej 17, DK-2100 Copenhagen \O, Denmark
}

\date{Received 7 March 2003 / Accepted 22 May 2003}
\offprints{T.\ A.\ Yousef, \email{tarek.yousef@mtf.ntnu.no}}

\abstract{
In the past few years suggestions have emerged that the solar magnetic
field might have a bi-helical contribution with oppositely polarized magnetic fields
at large and small scales, and that the shedding of such fields may
be crucial for the operation of the dynamo.
It is shown that, if a bi-helical field is shed into the solar wind,
positive and negative contributions of the magnetic helicity spectrum
tend to mix and decay.
Even in the absence of turbulence, mixing and decay
can occur on a time scale faster than the resistive one provided the two signs of
magnetic helicity originate from a single tube.
In the presence of turbulence, positively and negatively polarized
contributions mix rapidly in such a way that the ratio of magnetic
helicity to magnetic energy is largest both at the largest scale
and in the dissipation range.
In absolute units the small scale excess of helical fields is however negligible.
}

\maketitle

\newcommand{\comment}[1]{\footnote{#1}}

\section{Introduction}

Hydromagnetic turbulence in the solar wind is long known to have a
{
magnetic power spectrum that is compatible with $k^{-5/3}$
(although with substantial error bars, see Matthaeus et al.\ 1982).
}
The magnitude of the magnetic helicity is small, but there
is some evidence that it is negative above the
heliospheric current sheet and positive below it;
see Smith \& Bieber (1993) and Bieber \& Rust (1995).
Spectra of magnetic energy and helicity have later also been calculated by
Goldstein et al.\ (1994), Leamon et al.\ (1998), and Smith (1999) under
the assumption of isotropy.
They found net magnetic helicity only in the dissipation range.
They also reported efficient cancellation of
magnetic helicity in the inertial range.

The presence of magnetic helicity at small scales is puzzling, because
magnetic helicity is known to cascade to large scales
(Frisch et al.\ 1975, Pouquet et al.\ 1976).
This inverse cascade has also been seen in turbulence simulations both with helical
forcing (Pouquet \& Patterson 1978,
Balsara \& Pouquet 1999, Brandenburg 2001) and without forcing
{
but helical initial fields that are either tube-like
(Horiuchi \& Sato 1985, 1986, Zhu et al.\ 1995)
or turbulent (Christensson et al.\ 2001).
}
It is possible, however, that the spectral transfer of magnetic energy
from small to large scales is a process that does not happen on a
{
dynamical time scale but on a much slower
resistive time scale (Brandenburg \& Sarson 2002).
In that case the small scale magnetic helicity will not have had time to
cascade to larger scales in  the  interval between the time of ejection of the
observed magnetic field patch from the solar surface and the time of
in-situ observation.
}

The purpose of the present paper is to study the type
of interactions that can occur on short dynamical time scales.
This is important in connection with the solar magnetic field.
In active regions the field is found to be
markedly helical (Seehafer 1990, Pevtsov et al.\ 1995,
Rust \& Kumar 1996, Bao et al.\ 1999).
On the other hand,
the observed absence of net magnetic helicity in the inertial range of
the solar wind seems to be in conflict with { the observed markedly
helical field on the solar surface.}
We believe that
this may be related to recent theoretical suggestions that the magnetic
helicity ejected from the sun is {\it bi-helical} and has two components
with opposite magnetic helicity: one from intermediate scale fields and
one from the global field of the sun (Blackman \& Brandenburg 2003).

The bi-helical nature of the solar magnetic field is highlighted
by the fact that in active regions the magnetic helicity is negative
(and positive on the southern hemisphere) while, on the other hand,
bipolar regions
are tilted according to Joy's law (Hale et al.\ 1919),
i.e.\ in the clockwise direction in
the north (and anti-clockwise in the south) suggesting that the swirl of
flux tubes is right handed, giving rise to positive magnetic helicity
(and negative magnetic helicity in the south).
This corresponds to what is also known as {\it writhe} helicity
(e.g.\ Longcope \& Klapper 1997, D\'emoulin et al.\ 2002).
This idea of a bi-helical field is further supported by studies of sigmoids.
Figure~2a of Gibson et al.\ (2002) shows a TRACE image of an
{\sf N}-shaped sigmoid (right-handed writhe) with left-handed twisted filaments
of the active region NOAA AR 8668, typical of the northern hemisphere.

The puzzle of a bi-helical field structure can be resolved when one realizes
that an overall tilting of flux tubes causes simultaneously an equal amount
of internal twist in the tube (Blackman \& Brandenburg 2003).
This is best seen when the magnetic field is pictured as a ribbon.
This picture supports the notion that the magnetic field generated in
the sun by an $\alpha$-effect
must have opposite signs of magnetic helicity at large and
small scales (Seehafer 1996, Ji 1999, Brandenburg 2001,
Field \& Blackman 2002, Blackman \& Brandenburg 2002).

{
Once a magnetic field structure has emerged at the solar surface,
the tilt angle is known to relax gradually on a time scale of a few
days (Howard 1996, Longcope \& Choudhuri 2002).
This means that also the internal twist must decreases.
In order to understand quantitatively the fate of a bi-helical magnetic field
}
after it has been ejected into the solar wind we study,
using simulations of hydromagnetic turbulence, what happens to a magnetic
field that is composed of large and small scale fields of opposite
helicity and different amplitudes.
The expansion of the solar wind is not explicitly taken into account, but it
is known that under similar circumstances of cosmological expansion
helical magnetic fields can still show the inverse cascade effect
(Brandenburg et al.\ 1996).

\section{The model}

We consider a compressible isothermal gas with constant sound speed
$c_{\rm s}$, constant kinematic viscosity $\nu$, constant magnetic
diffusivity $\eta$, and constant magnetic permeability $\mu_0$. The
governing equations for density $\rho$, velocity $\uu$, and magnetic
vector potential $\AAA$, are given by
\EQ
{\DD\ln\rho\over\DD t}=-\nab\cdot\uu,
\EN
\EQ
{\DD\uu\over\DD t}=-c_{\rm s}^2\nab\ln\rho+{\JJ\times\BB\over\rho}
+\FF_{\rm visc}+\ff,
\label{dudt}
\EN
\EQ
{\partial\AAA\over\partial t}=\uu\times\BB-\eta\mu_0\JJ,
\label{dAdt}
\EN
where ${\rm D}/{\rm D}t=\partial/\partial t+\uu\cdot\nab$ is the
advective derivative, $\BB=\nab\times\AAA$ is the magnetic field,
and $\JJ=\nab\times\BB/\mu_0$ is the current density.
The viscous force is
\EQ
\FF_{\rm visc}=\nu\left(\nabla^2\uu+\onethird\nab\nab\cdot\uu
+2\nu\SSSS\cdot\nab\ln\rho\right),
\EN
where ${\sf S}_{ij}=\half(u_{i,j}+u_{j,i})-\onethird\delta_{ij}
\nab\cdot\uu$ is the traceless rate of strain tensor.
In cases with forcing, $\ff\neq0$ is a nonhelical forcing function,
selected randomly at each time step from a set of nonhelical transversal
wave vectors
\EQ
\ff_{\kk}=\left(\kk\times\eee\right)/\sqrt{\kk^2-(\kk\cdot\eee)^2},
\EN
where $\eee$ is an arbitrary unit vector needed in order to generate a
vector $\kk\times\eee$ that is perpendicular to $\kk$.

We use nondimensional quantities by measuring $\uu$ in units of $c_{\rm s}$,
$\xx$ in units of $1/k_1$, where $k_1$ is the smallest wave number in the box
(side length $L=2\pi$), density in units of the initial value $\rho_0$,
and $\BB$  is measured in units of $\sqrt{\mu_0\rho_0}\,c_{\rm s}$.
This is equivalent to putting $c_{\rm s}=k_1=\rho_0=\mu_0=1$.

{ 
In a periodic domain the total helicity, \EQ H=\int\AAA\cdot\BB\,\dd V,  \EN is
gauge invariant and conserved in the limit of zero magnetic diffusivity.  $H$ is a
topological measure of the mutual linkage of the magnetic flux lines and thus
the complexity of the magnetic field.  

Since we are interested in the distribution of helicity between different
scales it is convenient to decompose the total magnetic helicity into
spectral modes,  $H_k$, normalized such that
\EQ H =  \int H_k\,\dd k. \EN  $H_k$ is the contribution to
$H$ from the wave number interval (e.g.\ Brandenburg et al.\ 2002)
\EQ
k-\delta k/2 < |\kk| < k+\delta k/2\quad
\mbox{($k$-shell)}.
\EN
}

\section{Results}

\subsection{No forcing}

We first consider the evolution of a simple bi-helical initial field
consisting of a
superposition of two Beltrami waves with wave numbers $\kk_1=(0,0,1)$
and $\kk_5=(5,0,0)$ and helicities $H_1=10^{-6}$ and $H_5=-10^{-5}$.
The vector potential is then
\EQ
\AAA=\sqrt{H_1\over k_1}\pmatrix{\sin k_1 z\cr\cos k_1 z\cr0}
+\sqrt{-H_5\over k_5}\pmatrix{0\cr\cos k_5 x\cr\sin k_5 x}.
\EN
The net magnetic helicity is negative, but since the small scales decay
faster there will be a time when the net magnetic helicity turns positive.

\begin{figure}[t!]\centering\includegraphics[width=0.45\textwidth]{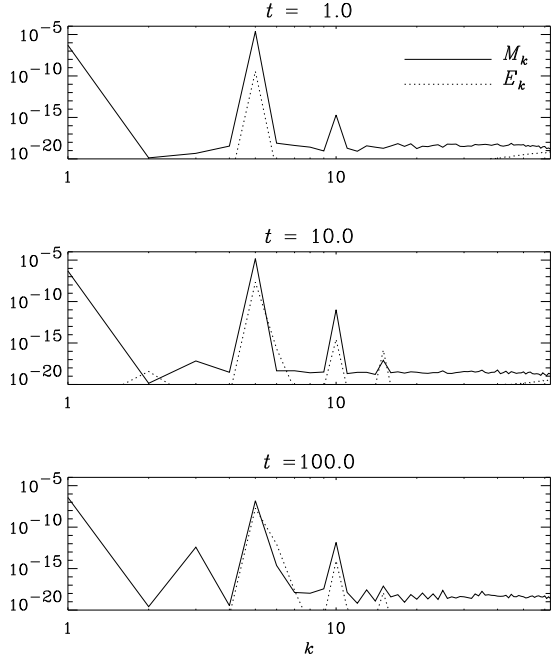}\caption{
Magnetic and kinetic power spectra  
at three different times.
No explicit forcing of the flow.
The initial magnetic field consists of a superposition of two
Beltrami fields at $\kk=(0,0,1)$ with $H_1=10^{-6}$ and
at $\kk=(5,0,0)$ with $H_5=-10^{-5}$.
}\label{Frelax128ef}\end{figure}

{
\FFig{Frelax128ef} shows magnetic and kinetic power spectra,
\EQ
E_k = \half\!\!\!\int\limits_{\mbox{$k$-shell}}
\!\!\!|\uu_{\kk}|^2\dd\kk,\quad
M_k = \half\!\!\!\int\limits_{\mbox{$k$-shell}}
\!\!\!|\BB_{\kk}|^2\dd\kk,\quad
\EN
at three different times.
}
We find (as expected) resistive decay of the power
on the two wave numbers.  
Thus, there is no enhanced decay.
This is readily explained by the absence of any spectral
overlap between the two components.
This would not change even if the wave vectors of the two Beltrami waves
were pointing in the same direction.

\subsection{Nonhelical forcing}

\begin{figure}[t!]\centering\includegraphics[width=0.45\textwidth]{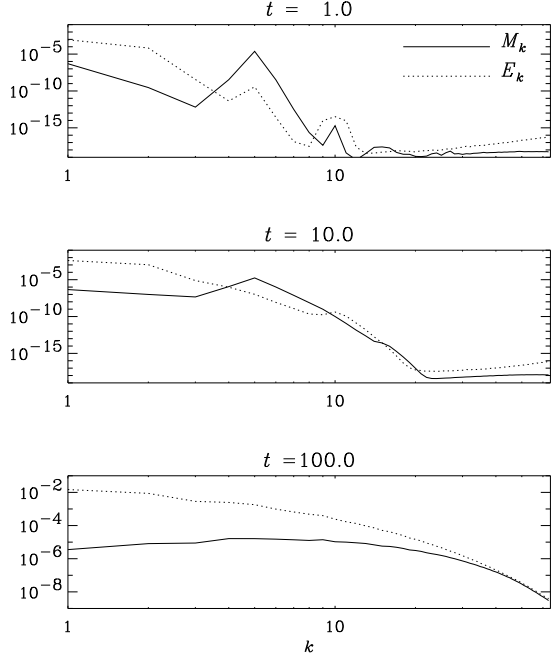}\caption{
Magnetic and kinetic power spectra at three different times.
The flow is forced nonhelically
{
in the wave numbers interval $1\leq|\kk|\leq2$}.
The initial magnetic field consists of a superposition of two
Beltrami fields at $\kk=(0,0,1)$ with $H_1=10^{-6}$ and
at $\kk=(5,0,0)$ with $H_5=-10^{-5}$.
}\label{Frelax128ek}\end{figure}

We now drive turbulence by adding a random forcing term that acts
on large scales with wave numbers between 1 and 2.
We find that the magnetic power in the two modes at $k=1$ and 5
spreads rapidly among other wave numbers; see \Fig{Frelax128ek}.
In addition, there is also dynamo action giving rise to a magnetic power
spectrum that rises with $t$; see the last panel of \Fig{Frelax128ek}.
The form of the magnetic energy spectrum in the dynamo case is
subject of a separate investigation (Haugen et al.\ 2003,
and references therein).

\begin{figure}[t!]\centering\includegraphics[width=0.45\textwidth]{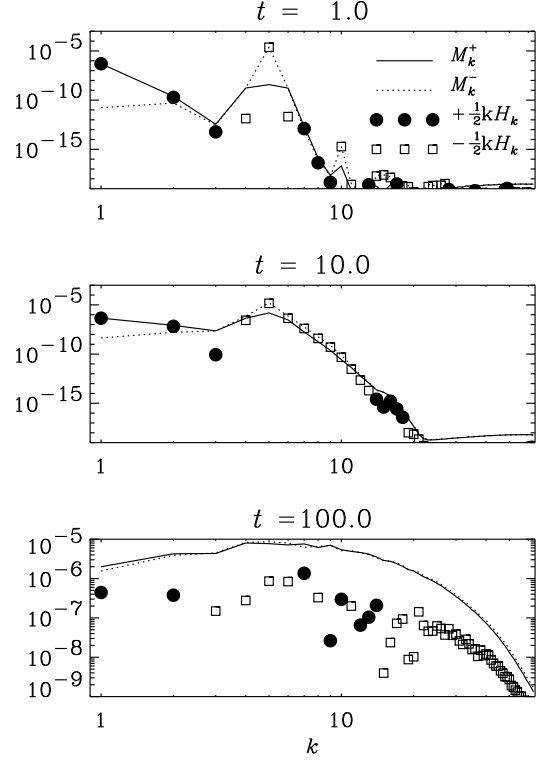}\caption{
Magnetic power spectra of positive and negative polarized
contributions to the magnetic field.
The sign of the magnetic helicity is indicated by
black dots (positive) and open squares (negative).
Otherwise like in \Fig{Frelax128ek}.
}\label{Frelax128ek_hel}\end{figure}

It is convenient to divide the magnetic power spectrum into
contributions from positively and negatively polarized waves such that
$M_k=M_k^++M_k^-$ (Waleffe 1993, see also Brandenburg et al.\ 2002).
The magnetic helicity spectrum can then be written as
\EQ
H_k=(2/k)(M_k^+-M_k^-).
\EN
The result is shown in \Fig{Frelax128ek_hel}, where we see that at early
times ($t\leq10$), $M_k^+$ peaks at $k=1$ and $M_k^-$ peaks at $k=5$.
The magnetic helicity spectrum, normalized by $k/2$, is positive for
$k\leq3$ and negative for $k$ around 5.
At later times, $M_k^+\approx M_k^-$ and so the normalized magnetic
helicity, $\half k H_k$, is small, suggesting that contributions from
positively and negatively polarized waves have mixed almost completely.

\begin{figure}[t!]\centering\includegraphics[width=0.45\textwidth]{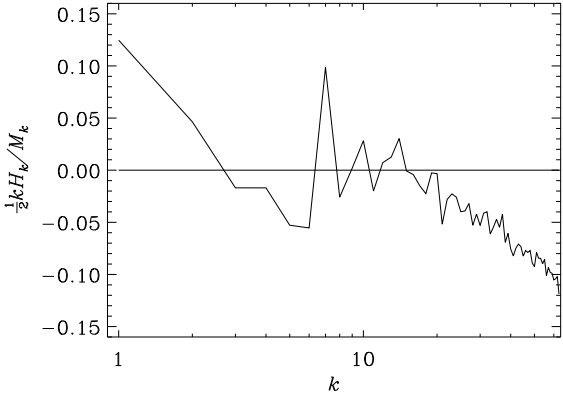}\caption{
Magnetic helicity spectrum normalized by magnetic energy
for the same run as shown in \Fig{Frelax128ek}, at $t=100$.
Note the relative magnetic helicity excess both at largest and smallest
scales.
}\label{Frelax128ek_hel2}\end{figure}

In \Fig{Frelax128ek_hel2} we plot $\half k H_k/M_k$, i.e.\ the magnetic
helicity spectrum relative to the magnetic energy spectrum.
Note the systematically negative values of the normalized magnetic
helicity in the dissipative subrange, just like in the plots
of Leamon et al.\ (1998).
This enhancement of (negative) magnetic helicity is probably
the result of a direct cascade that transfers the negative
magnetic helicity from intermediate to still smaller scales.
On the other hand, in absolute units the magnetic helicity remains small
at small scales and therefore the enhancement relative to the magnetic
energy spectrum could be regarded as a typical feature of such normalized plots
and is not an indication of a real excess of small scale magnetic helicity.

To summarize, the decay of magnetic helicity by mixing occurs on a
resistive time scale if there is no externally driven turbulence.
Decay on a faster (turbulent) time scale is possible in the presence of
externally driven nonhelical turbulence.
This is also in qualitative agreement with simulations by
Maron \& Blackman (2002) who found that if the helicity of the forcing
is below a certain threshold the large scale field decays rapidly.
In the remainder of this paper we consider the decay of bi-helical fields
without any driving.

\subsection{Decay of a random bi-helical field}

\begin{figure}[t!]\centering\includegraphics[width=0.45\textwidth]{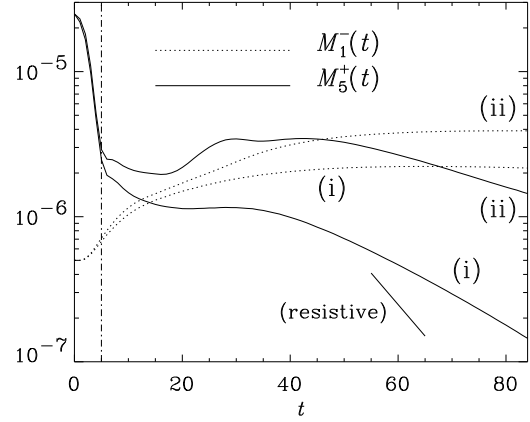}\caption{
Decay of $M_k^\pm(t)$ for $k=1$ (dotted lines) and $k=5$ (solid lines).
The lower curves, denoted by (i), are for $\nu=2\times\eta=10^{-3}$
while the upper ones, denoted by (ii), are for $\nu=\eta=10^{-3}$.
The run that was forced helically until $t=5$, and was then let to decay.
Note that the field never decays faster than on the resistive rate
(here only indicated for $k=5$).
}\label{Fphel_decay_relax64l}\end{figure}

We first consider a random initial condition obtained by stopping a
helically driven turbulence run (as studied in Brandenburg 2001).
In \Fig{Fphel_decay_relax64l} we show the evolution of $M_k^+$ and
$M_k^-$ for those values of $k$ where the spectra are maximum for
the initial condition ($k=1$ for $M_k^-$ and $k=5$ for $M_k^+$).
Note that both $M_k^+$ and $M_k^-$ decay at rates that are clearly
slower than the resistive rate (indicated by a straight line).
This is because in three dimensional turbulence the decay of a magnetic
field is easily offset by dynamo action, even if the velocity is decaying
(Dobler et al.\ 2003).

In addition to the field decaying only slowly we also find that the
bi-helical character of the field prevails.
Thus, dynamo action alone does not produce mixing of the positively and
negatively polarized components.
In the following section we suggest that this is primarily a shortcoming
of our initial condition lacking positively and negatively polarized
components within a single physically connected flux structure.

\subsection{Decay of a bi-helical flux tube}
\label{bihel_tube}

In order to make the connection with the magnetic field structure in
real space it is instructive to consider explicitly a twisted flux tube.
We do this here by constructing the Cauchy solution of an initially
straight flux tube in a simple steady flow field.
The Cauchy solution (e.g.\ Moffatt 1978) is
\EQ
B_i(\xx,\tau)=
{{\sf G}_{ij}(\xx_0,\tau)\over\det\GGGG}
\,B_{0j}(\xx_0),
\EN
where $\tau$ is the time coordinate for constructing the solution,
${\sf G}_{ij}=\partial x_i/\partial x_{0j}$ is the Lagrangian
displacement matrix, $\BB_0(\xx_0)=\BB(\xx,0)$ is the initial condition, and
\EQ
\xx(\xx_0,\tau)=\xx_0+\int_0^\tau\UU\left(\xx(\tau')\right)\dd\tau'
\label{lagrange}
\EN
is the position of an advected test particle whose original position
was at $\xx_0$.

\begin{figure}[t!]\centering\includegraphics[width=0.45\textwidth]{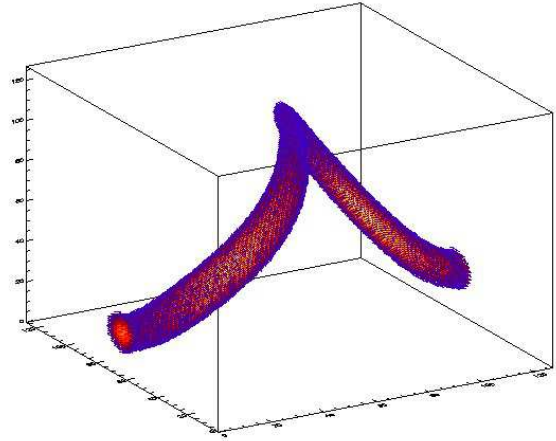}\caption{
Magnetic flux tube constructed from the Cauchy solution.
$\tau=2$, $\varphi=0.2$.
}\label{Fvecgd_t0}\end{figure}

\begin{figure}[t!]\centering\includegraphics[width=0.45\textwidth]{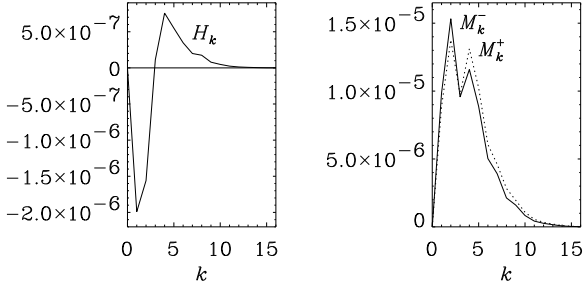}\caption{
Initial spectra of magnetic helicity, $H_k$, and of magnetic energy
of positively and negatively polarized components, $M_k^+$ and $M_k^-$,
respectively.
}\label{Fpower_init}\end{figure}

The flow must have the property of lifting the tube only in one section
and tilting it sideways, hence causing writhe helicity in the tube.
A simple steady flow with such properties is
\EQ
\UU(\xx)=\pmatrix{0\cr \varphi z\sin x\cr1+\cos x},
\EN
where $\varphi$ is a parameter that controls the amount of twist.
Periodic boundary conditions are assumed in the range
$-\pi<(x,y)<\pi$ and $0<z<2\pi$.
With this velocity field, \Eq{lagrange} reduces to
\EQ
\xx(\xx_0,\tau)=\xx_0+\UU(\xx_0)\,\tau.
\EN
This yields
\EQ
{\sf G}_{ij}(\xx_0,\tau)
=\pmatrix{1 & 0 & 0\cr
z_0\varphi\tau\cos x_0 & 1 & \varphi\tau\sin x_0\cr
-\tau\sin x_0 & 0 & 1}.
\EN
Note that $\det\GGGG=1$ (incompressibility) and
${\sf G}_{ij}(\xx_0,0)=\delta_{ij}$, so $\BB_0(\xx_0)=\BB(\xx,0)$.
An example of the resulting magnetic field structure is shown in \Fig{Fvecgd_t0}
for $\varphi=0.2$ and $\tau=2$.
The initial field was a straight tube with
\EQ
\BB_0(\xx)=\xxx B_0\exp\{-[y^2+(z-h)^2]/d^2\},
\EN
where $h=1$ is the initial height of the tube,
$d=0.5$ is its radius, and $B_0=0.1$ is the initial field strength.
The magnetic helicity spectrum shows distinctively bi-helical
behavior; see \Fig{Fpower_init}.

\begin{figure}[t!]\centering\includegraphics[width=0.45\textwidth]{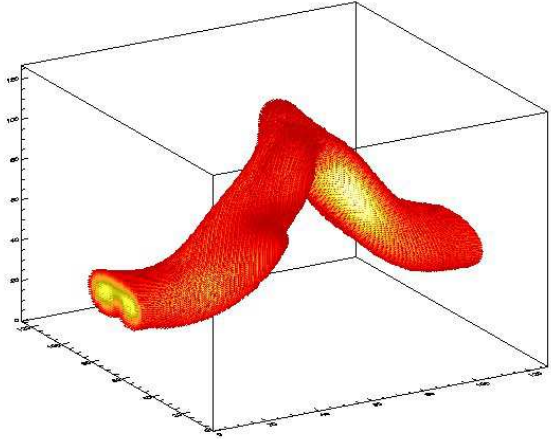}\caption{
Magnetic flux tube after letting it relax until $t=33$.
}\label{Fvecgd_t33}\end{figure}

\begin{figure}[t!]\centering\includegraphics[width=0.45\textwidth]{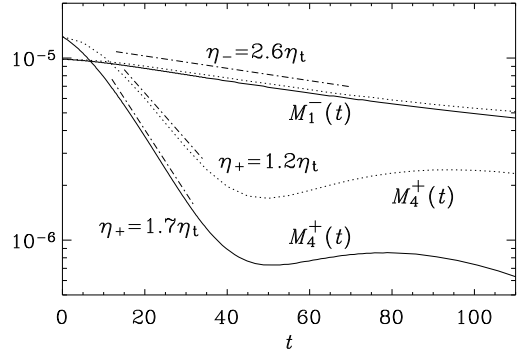}\caption{
Decay of $M_1^-(t)$ and $M_4^+(t)$ for $\nu=\eta=10^{-3}$ { (solid lines)}.
The dotted lines give the result for $\nu=10^{-3}$ and $\eta=10^{-4}$.
The slopes 
{ (dash-dotted lines)}
are given in units of $\eta_{\rm t}\equiv u_{\rm rms}/k_{\rm eff}$,
where $u_{\rm rms}$ is the peak rms velocity and $k_{\rm eff}=4$ is the
estimated effective wave number of the energy carrying scale during
the time where the velocity is maximum.
The upturn in the $M_4^+(t)$ curves after $t=40$ is temporary and caused
by field cascading backward from larger wave numbers.
}\label{Fphel_decay}\end{figure}

We use this Cauchy solution as initial condition for
our simulation, i.e.\ $t=0$ corresponds now to $\tau=2$.
The result at time $t=33$ is shown in \Fig{Fvecgd_t33}.
Unlike the previous case (\Figs{Frelax128ef}{Fphel_decay_relax64l}), the
magnetic energies of the positively and negatively polarized components
decay at a rate faster than the resistive rate.
We characterize the decay of $M_\pm(k_\pm,t)$ at $k_-=1$ and $k_+=4$
in terms of effective diffusivity coefficients, $\eta_\pm$, by writing
the decay law as
\EQ
M^\pm(k_\pm,t)\sim\exp(-2\eta_\pm k_\pm^2 t).
\EN
The result is shown in \Fig{Fphel_decay} for two different values of $\eta$.
The decay rates are either completely independent of the microscopic value
of $\eta$ (for $k=1$), or depend on it only weakly (for $k=4$).
The decay happens therefore on something like a `turbulent' resistive
time scale.
For orientation we use the maximum rms velocity $u_{\rm rms}\approx0.006$
and the estimated effective wave number of the energy carrying scale
during velocity maximum, $k_{\rm eff}=4$, to calculate a `turbulent'
magnetic diffusivity, $\eta_{\rm t}\equiv u_{\rm rms}/k_{\rm eff}$.
It turns out that the effective diffusivity coefficients, $\eta_\pm$, are
{
of the order of $\eta_{\rm t}$ and independent of the magnetic Reynolds
number. A relevant measure of the magnetic Reynolds is $\eta_{\rm t}/\eta$.
This ratio is either 1.5 or 15 in the two cases shown (\Fig{Fphel_decay},
solid and dashed lines, respectively), so the magnetic Reynolds
number has changed by a factor of 10 while the actual decay rates
(characterized by the values of $\eta_+$ and $\eta_-$) are almost
unchanged (\Fig{Fphel_decay}).
}

We conclude from this that a single flux tube in a bi-helical state with
writhe and twist helicity of opposite sign relaxes to a nonhelical state
on a dynamical time scale.
The final state is not, however, a simple straight tube, but one with a
more complicated internal structure.
On the other hand, bi-helical fields generated randomly decay more slowly
and do not tend to mix.
This is explained by dynamo action of the decaying turbulent flow field
and by inverse transfer of magnetic helicity (Pouquet et al.\ 1976,
Christensson et al.\ 2001).

\section{Discussion}

Helical driving in the solar convection zone tends to produce bi-helical
fields.
In the northern hemisphere the magnetic helicity is probably positive
at large scales and negative at smaller scales.
In the absence of helical driving, for example in the
solar wind, an initially bi-helical field can relax in such a way that
contributions with positive and negative magnetic helicity mix.
The decay of magnetic helicity in the different modes will in general
happen on a turbulent time scale, until a certain level is reached where
a nonhelical magnetic field can be sustained by nonhelical dynamo action.
Even in the absence of any driving magnetic helicity of opposite
signs and at different scales can mix, provided this involves a
singly connected magnetic field structure (as shown in \Sec{bihel_tube}).
This is probably related to the possibility of rapid propagation of twist
along the tube as a torsional Alfv\'en wave (Longcope \& Klapper 1997).
If the different signs of helicity do not reside on a single tube,
the mixing of oppositely oriented writhe and twist can be as slow as a resistive
time scale, as is the case where two orthogonal Beltrami waves with opposite
sign and different wavelengths are superimposed.

In order to see the distribution of the sign of the magnetic helicity
over different scales one often plots the magnetic helicity spectrum
normalized by the magnetic energy spectrum.
For the solar wind above the heliospheric current sheet this ratio fluctuates
around zero and has a negative excess within the dissipative subrange
(Leamon et al.\ 1998).
Similar behavior is also found in the present case and seems to be
indeed a consequence of magnetic diffusion helping mix oppositely oriented waves
efficiently.

Although there is qualitative support of the idea that there is large
scale magnetic helicity with opposite sign (Gibson et al.\ 2002), there
is to our knowledge no quantitative measurement.
All explicit measurements point to a negative sign in the northern hemisphere.
If the solar field is indeed bi-helical, there may be as yet undetected
positive magnetic helicity at larger or perhaps even at smaller scales.
The fact that in the dissipative range of solar wind turbulence
the magnetic helicity
is negative supports the former suggestion that the negative
helicity seen at the solar surface is related to the small scale magnetic
helicity, and that there is positive magnetic helicity on scales larger
than the scale of active regions.

Ideally, one should try to measure this large scale magnetic helicity from the
longitudinally averaged mean field.
 From synoptic maps one can obtain the longitudinally averaged radial
field at the surface.
This allows one to calculate the toroidal component of the magnetic
vector potential, $\meanA_\phi$.
Using spherical polar coordinates for the axisymmetric mean field, the
gauge-invariant magnetic helicity of Berger \& Field (1984) is simply
$H=2\int\meanA_\phi\meanB_\phi\,\dd V$ (see Brandenburg et al.\ 2002).
In order to calculate $H$ one still needs $\meanB_\phi$.
Unfortunately, this cannot be observed directly.
Only its sign can be inferred from the orientation of bipolar regions.
Preliminary investigations (Brandenburg et al.\ 2003) suggest
that there are approximately equally long intervals where
$\meanA_\phi\meanB_\phi$ is positive and negative during one cycle.
The result is therefore inconclusive.

The best support for oppositely oriented magnetic helicity
at large scales comes from the fact that the footpoints of emerging
magnetic flux tubes are tilted clockwise in the northern hemisphere and
counterclockwise in the southern hemisphere (Joy's law).
This corresponds to right-handed tubes in the northern hemisphere and
left-handed tubes in the southern hemisphere, i.e.\ to positive magnetic
helicity in the north and negative magnetic helicity in the south.
This is also in agreement with studies of sigmoids suggesting that on
the northern hemisphere {\sf N}-shaped sigmoids with right-handed writhe
have left-handed twisted filaments (Gibson et al.\ 2002).

\begin{acknowledgements}
We thank Eric Blackman and Annick Pouquet for comments on the paper.
Use of the supercomputers in Odense (Horseshoe), Trondheim (Gridur),
and Leicester (Ukaff) is acknowledged.
\end{acknowledgements}


\end{document}